\documentclass[12pt]{article}
\usepackage{amssymb,amsmath,amsthm,graphicx,ulem}

\pdfoutput=1

\usepackage{graphicx,subfigure}
\usepackage{epsfig}
\usepackage{amsmath}
\usepackage{amsfonts}
\usepackage{amssymb}
\usepackage[usenames]{color}
\usepackage[letterpaper,left=2.cm,right=2.cm,top=2.5cm,bottom=2.5cm]{geometry}

\newcommand{\beq}{\begin{equation}}
\newcommand{\eeq}{\end{equation}}
\newcommand{\be}{\begin{equation}}
\newcommand{\ee}{\end{equation}}
\newcommand{\beqa}{\begin{eqnarray}}
\newcommand{\eeqa}{\end{eqnarray}}
\newcommand{\beqar}{\begin{eqnarray*}}
\newcommand{\eeqar}{\end{eqnarray*}}
\newcommand{\bea}{\begin{eqnarray}}
\newcommand{\eea}{\end{eqnarray}}

\newcommand{\ie}{{\it i.e.}\ }

\numberwithin{equation}{section}

\newcommand{\nn}\nonumber
\newcommand{\eqn}[1]{(\ref{#1})}

\numberwithin{equation}{section}

\begin{document}

\allowdisplaybreaks

\normalem

\title{Further Evidence for  Lattice-Induced Scaling}

\author{Gary T. Horowitz${}^{\,a}$,
Jorge E. Santos${}^{\,a}$, David Tong${}^{\,b}$\\ 
\\ \\
  ${}^{\,a}$ Department of Physics, UCSB, Santa Barbara, CA 93106, USA \\ 
  ${}^{\,b}$ DAMTP,  University of Cambridge, Cambridge, CB3 0WA, UK \\
 \\ 
 \small{gary@physics.ucsb.edu, jss55@physics.ucsb.edu, d.tong@damtp.cam.ac.uk}}

 \date{}

\maketitle

\begin{abstract}
\noindent  We continue our study of holographic transport in the presence of a background lattice.  We recently found evidence that the presence of a lattice induces a new intermediate scaling regime in asymptotically $AdS_4$ spacetimes. This manifests itself in the optical conductivity which exhibits a robust power-law dependence on frequency, $\sigma \sim \omega^{-2/3}$, in a ``mid-infrared" regime, a result which is in striking agreement with experiments on the cuprates. Here we provide further evidence for the existence of this intermediate scaling regime. We demonstrate similar scaling in the thermoelectric conductivity,   find analogous scalings in asymptotically $AdS_5$ spacetimes, and 
 show that we get the same results with an ionic lattice. 
 \end{abstract}

\newpage


\tableofcontents
\baselineskip16pt

\section{Introduction}
As holographic methods develop as a tool to understand strongly correlated condensed matter systems, it has become increasingly important to explore gravitational backgrounds  which are not spatially homogeneous. Recent examples include the study of physics in the presence of a background lattice \cite{us,seandiego,maeda,maeda2,schalm,aristos}, lattices constructed from D-branes \cite{shamit1,shamit3}  as well as spatially modulated phases in which the breaking of translational invariance occurs spontaneously \cite{ooguri,ooguri2,monopole,jerome1,jerome}.

One arena in which the presence or absence of a lattice can dramatically affect the physics is transport at finite density. In 
 translationally invariant systems, momentum is necessarily conserved which, in turn, means that there can be no mechanism to dissipate zero frequency currents. This results, for example, in a delta-function peak in the optical conductivity  at zero frequency.

In a recent paper \cite{us}, we studied the optical conductivity in the presence of a lattice imposed as a spatially-varying source  on a $d=2+1$ dimensional boundary theory. This lattice imprints itself as ripples in the  bulk Reissner-Nordstr\"om $AdS_4$  spacetime and perturbations of  this geometry provide  a holographic computation of the optical conductivity in a more realistic setting. 
 The delta-function at $\omega=0$ is resolved, revealing  a number of interesting features lying beneath. 
 The most surprising result of \cite{us} occurred in an intermediate frequency range.  
 Here the optical conductivity exhibits power-law behaviour with
\be |\sigma(\omega)| = \frac{B}{\omega^{\gamma}}+C\label{power}\ee
To within numerical accuracy, $\gamma = 2/3$. 

Strikingly, the exponent $\gamma = 2/3$  is in agreement with measurements of bismuth-based  cuprates which also exhibit a power-law optical conductivity  in the same regime  \cite{vandermarel,marel2}. (See also \cite{azrak} for earlier experimental results and \cite{anderson,kato,norman} for attempts to explain this scaling). Also in agreement with the measurements is the fact that the coefficient $B$ is temperature independent. However, the cuprates do not appear to have the off-set $C$.

The results of \cite{us} strongly suggest the emergence of a new scaling region induced by the presence of the lattice. This is not obviously related to either the ultra-violet $AdS_4$ geometry, nor the infra-red $AdS_2\times {\bf R}^2$ geometry, but appears in an intermediate energy regime. At present, we have little understanding of how this scaling arises and in this note we do not shed much light on the issue. Instead, we merely offer further evidence for the existence of this scaling region. We present three new results. First, we compute the thermoelectric conductivity in the presence of the lattice and find that it too exhibits a power-law scaling in the same regime. Second, we repeat the calculation of optical conductivity in asymptotically $AdS_5$ spacetimes and find that there is again a scaling regime \eqn{power}, albeit with a different exponent: $\gamma \approx \sqrt 3/2$. Finally, we return to $AdS_4$ and construct an entirely different bulk spacetime lattice, now built from a spatially varying chemical potential. We  show that the power-law behavior \eqn{power} persists with the same exponent $\gamma = 2/3$. We also find resonances in the optical conductivity similar to those seen in many condensed matter systems. We argue that such resonances are a generic feature of holographic lattices.

\section{\label{sec:model}Review of Earlier Results}
We begin by summarizing the set up and results of \cite{us}.  We consider the minimal ingredients necessary to compute conductivity in a holographic framework, namely Einstein-Maxwell theory in $AdS_4$.  To this we add a neutral scalar field $\Phi$ which we  use to source the lattice, so we work with the action
\begin{equation}
S= \frac{1}{16 \pi G_N}\int d^4 x\,\sqrt{-g}\left[R+\frac{6}{L^2}-\frac{1}{2}F_{ab}F^{ab}-2\nabla_a \Phi \nabla^a \Phi +\frac{4\Phi^2}{L^2}\right],
\label{eq:action}
\end{equation}
where  $L$ is the AdS length scale and $F= \mathrm{d} A$. 
The scalar mass, $m^2 = -2/L^2$, is chosen  since for this choice, the asymptotic behavior of $\Phi$ is simple. If the metric asymptotically takes the standard form
\be ds^2 = \frac{-dt^2 + dx^2 + dy^2 + dz^2}{z^2}
\ee
then
\be \Phi = z\phi_1 + z^2\phi_2 + {\cal O}(z^3)\ee

We  introduce the  lattice  by providing a spatially inhomogeneous boundary condition for $\Phi$,  \ie by specifying  $\phi_1(x,y)$.  For numerical convenience, we only introduce the lattice in one direction and choose   $\phi_1$ to be 
\begin{equation}
\phi_1(x) = A_0 \cos (k_0 x)\,.
\label{eq:source1}
\end{equation}
The lattice size $l$ is simply the reciprocal of the lattice wavenumber: $l =2\pi/k_0$. 
This corresponds to adding a periodic source to the  dimension two operator dual to $\Phi$. Since $\phi_1$ is  translationally invariant in the $y$ direction, it is perhaps better to describe this system as striped rather than as  a lattice. This simplifies the numerics  since the resulting PDEs are only two-dimensional rather than three-dimensional. 

Since we want to work at finite temperature and finite density, we add an electrically charged black hole in the interior.  This requires a nonzero vector potential $A_t$ 
which vanishes at the horizon and is asymptotically
\be\label{asympAt}
A_t = \mu - \tilde\rho(x) z + O(z^2)
\ee
where $\mu$ is a constant chemical potential and $\tilde\rho(x)$ is the charge density.
We numerically solve the coupled Einstein-Maxwell-scalar equations, subject to the boundary conditions \eqn{eq:source1} and \eqn{asympAt} using the Einstein-DeTurck method described in \cite{Headrick:2009pv}. With five metric functions together with $A_t$ and $\Phi$, there are a total of seven coupled PDE's to solve. Full details of the metric ansatz, boundary conditions on metric components and further explanation of numerical techniques can be found in \cite{us}. 

The resulting static solutions  represent our holographic lattice. The backreaction of the scalar field causes the metric and hence the Maxwell field to both become periodic in the $x$ direction. They  are effectively rippled Reissner-Nordstr\"om-AdS black holes, parameterized by four variables: the  chemical potential $\mu$, the size of the modulation $A_0$, the lattice wavenumber $k_0$ and the temperature $T$. The resulting physics depends only on three independent ratios of these variables which we take to be $T/{\mu}$, $k_0/{\mu}$, and $A_0/k_0$.

To compute the optical conductivity in the direction of our lattice (against the grain of the stripes), we  perturb the background fixing the usual boundary condition on $\delta A_x$: 
\be\label{Abdycond}
 \delta A_x \rightarrow \frac{E}{i\omega} + J_x(x, \omega) z + O(z^2)
\ee
This corresponds to adding a homogeneous electric field $E_x = E e^{-i\omega t}$  on the boundary. 
 At linear order, this further  sources perturbations for the gauge field components $\delta A_t$ and $\delta A_z$, the scalar $\delta \Phi$, as well as the metric components $\delta g_{tt}$, $\delta g_{tz}$, $\delta g_{tx}$, $\delta g_{zz}$, $\delta g_{zx}$, $\delta g_{xx}$ and $\delta g_{yy}$. In other words, almost everything. This results in eleven coupled partial differential equations in two variables, $x$ and $z$. Solving these linear equations with suitable boundary conditions allows us to read off the current $J_x(x, \omega)$ which determines the conductivity via Ohm's law.  We can express the conductivity in a manifestly gauge invariant form as
 \begin{equation}
\tilde{\sigma}(\omega,x)\equiv\lim_{z\to 0}\frac{\delta F_{zx}(x,z)}{\delta F_{xt}(x,z)}.
\label{eq:opticalderiv}
\end{equation}
 Since we  impose a homogeneous electric field, we are interested in the homogeneous part of the conductivity, $\sigma(\omega)$.
As expected, the lattice resolves the zero frequency delta-function yielding the following results:
 \begin{itemize} 
\item The DC resistivity is finite and exhibits a dependence on temperature, $\rho \sim T^{2\Delta}$, where $\Delta$ is a (known) function of the lattice spacing. This property was previously predicted in \cite{seandiego} to occur in holographic models and can be traced to  the locally critical (i.e. with dynamical exponent $z\rightarrow \infty$)  $AdS_2$ near horizon geometry of the black hole.
\item At low frequencies (which, when $k_0$ and $A_0$ are the same of order magnitude as $\mu$,  can roughly be characterised as $\omega \lesssim T$) the optical conductivity follows the familiar Drude form
\be\label{drude}
\sigma(\omega) = \frac{K\tau}{1-i\omega \tau}
\ee
 with the scattering time $\tau$ temperature dependent but not frequency dependent. 
%
\item As mentioned in the introduction, the most interesting result of \cite{us} occurred in an intermediate frequency range, characterized by  $2<\omega \tau <8$.  Here the optical conductivity exhibits the power-law behaviour \eqn{power} with  $\gamma= 2/3$. 
\end{itemize}

\begin{figure}
\centerline{
\includegraphics[width=0.9\textwidth]{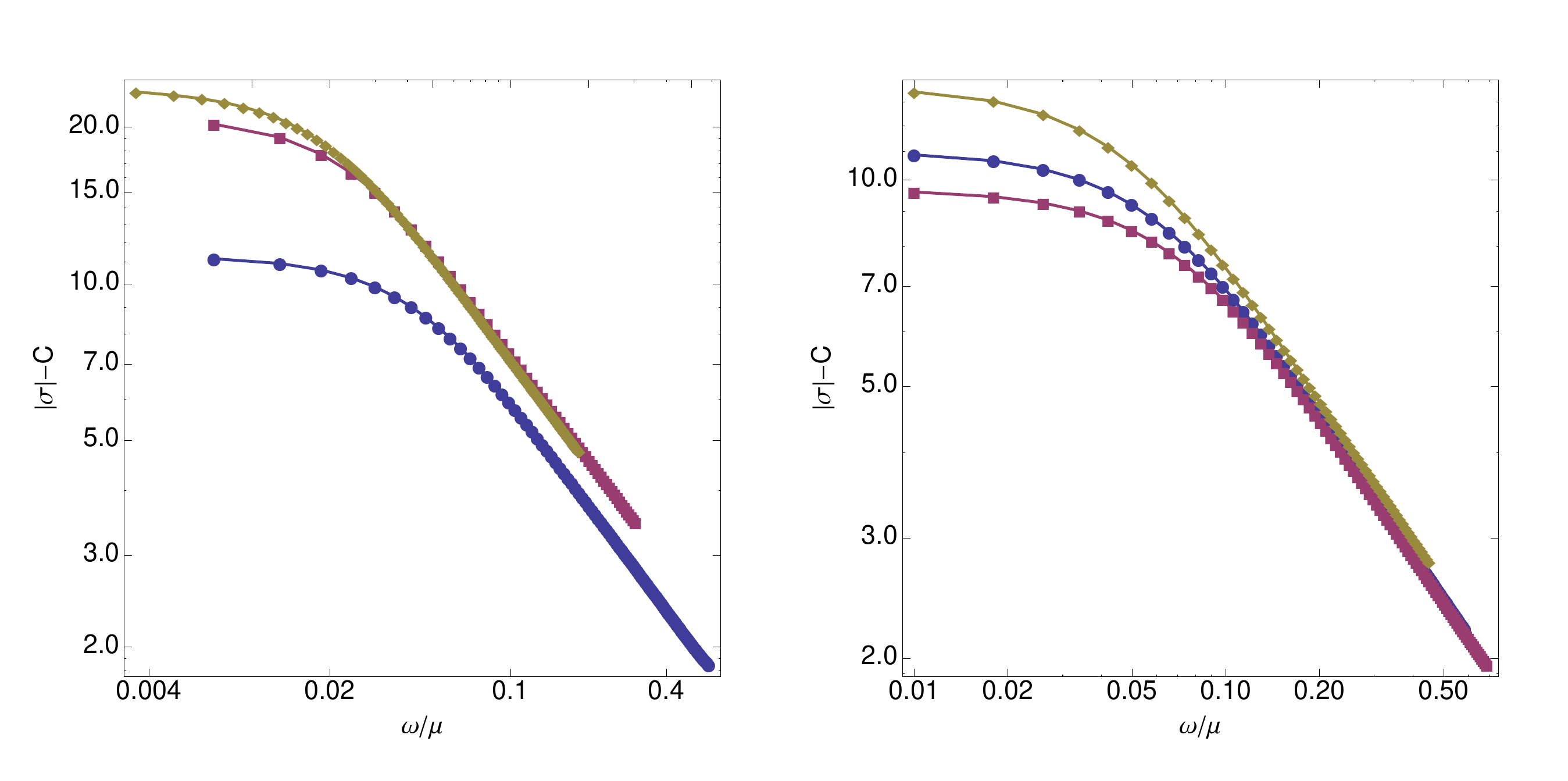}
}
\caption{A log-log plot of the optical conductivity for a $2+1$ dimensional system as a function of frequency. On the left, the plot has $T/\mu = .115$ and shows three different wavenumbers:  squares denote $k_0 = 1$, circles denote $k_0 = 2$, and diamonds denote $k_0 = 3$. The plot on the right has $k_0 = 2$ and shows three different temperatures: the diamonds have $T/\mu = .098$, the circles have $T/\mu = .115$ and the squares have $T/\mu = .13$. In both plots, $\mu = 1.4$ and $A_0/k_0 = 3/4$.}
\label{fig:4dlogs}
\end{figure}

The exponent $2/3$ is robust under changing the parameters in our model. Fig.~\ref{fig:4dlogs} shows the magnitude of the optical conductivity with the offset removed in a log-log plot for various $k_0$ and $T$.  This plot shows the same data as in Fig. 9 of \cite{us}, but we now plot $|\sigma| - C$ as a function of $\omega/\mu$ rather than $\omega\tau$. When plotted in terms of $\omega\tau$, it is clear that the scaling region is $2<\omega \tau <8$ in all cases. But since $\tau$ depends on $T$ and $k_0$,  to compare the amplitudes of the curves it is more accurate to use $\omega/\mu$. The fact that the lines on the left become parallel shows that the exponent is independent\footnote{We have also decreased $k_0$ to $1/2$ and found no change in the exponent.} of $k_0$, and the fact that they lie on top of each other on the right shows that not only is the exponent independent of $T$, the coefficient $B$ is also independent of temperature. This striking fact was obscured in \cite{us} since the data was plotted in terms of $\omega\tau$.  

\section{Thermoelectric Conductivity}\label{sec:alpha}

The power-law behaviour of the conductivity  indicates the presence of a new scaling regime induced by the presence of the lattice. It is clearly of interest to see if this also shows up in other transport properties. To this end, we extend our analysis to include both the electric current $J_x$ and the thermal current $Q_x={T^t}_{x}- \mu J_x$,
\be
 \left(\begin{array}{c} J_x \\ Q_x\end{array}\right) = \left(\begin{array}{cc} \sigma & \alpha  \\ \alpha T & \bar{\kappa} \end{array}\right)\left(\begin{array}{c} E_x \\ -\partial_x T
\end{array}\right)\nn\ee
where $\bar{\kappa}$ is the thermal conductivity and $\alpha$  the thermoelectric conductivity.

The holographic computation of these conductivities is standard fare: see, for example, \cite{herzog}. Since we have added a source for the scalar field,  the boundary stress tensor is not simply given by the $O(z^3)$ corrections to the metric. There is a contribution from the scalar field  as well.  Following \cite{Balasubramanian:1999re}, the stress energy tensor is  given by
\begin{equation}
T_{\mu\nu} = \lim_{z\to 0}\left(\frac{L}{z}\right)\left(K_{\mu\nu}-\gamma_{\mu\nu}K-\frac{2}{L}\gamma_{\mu\nu}+L G^{(3)}_{\mu\nu}-\frac{\Phi^2}{L}\gamma_{\mu\nu}\right)\,,
\label{eq:stressenergy}
\end{equation}
where Greek indices run over boundary coordinates, $K_{\mu\nu}$ is the extrinsic curvature associated with an inward unit normal vector to the boundary (located at $z=0$), $K\equiv \gamma^{\mu\nu}K_{\mu\nu}$, $\gamma_{\mu\nu}$ is the induced metric on the constant $z$ surface, and $G^{(3)}_{\mu\nu}$ is the Einstein tensor of $\gamma_{\mu\nu}$. Since we are interested in field theories living on Minkowski space, the fourth term in Eq.~(\ref{eq:stressenergy}) vanishes as $z\to0$. The last term, on the other hand, gives the necessary contribution to cancel the divergences arising due to the presence of the scalar field.

It is a numerical challenge to compute $T^{tx}$ since it  requires an accurate determination of  $O(z^3)$ corrections to the metric perturbation. For the thermal conductivity, this perturbation must have an order one contribution which describes the temperature gradient. Since we are perturbing off a numerical solution, we have not been able to reliably extract $T^{tx}$ in this case, and hence have not calculated the thermal conductivity $\bar{\kappa}$.

However, the thermoelectric conductivity  $\alpha$ is easier to compute since we only need to impose a boundary electric field as in (\ref{Abdycond}), and there is no order one contribution to the metric perturbation. We have computed the thermal current $Q_x $ in this case and determined  $\alpha = Q_x/TE_x$. The  results are shown in Fig.~\ref{fig:alphasmall} and Fig.~\ref{fig:alphapower}. 
The thermoelectric conductivity  $\alpha$ exhibits the following properties:

\begin{figure}
\centerline{
\includegraphics[width=0.9\textwidth]{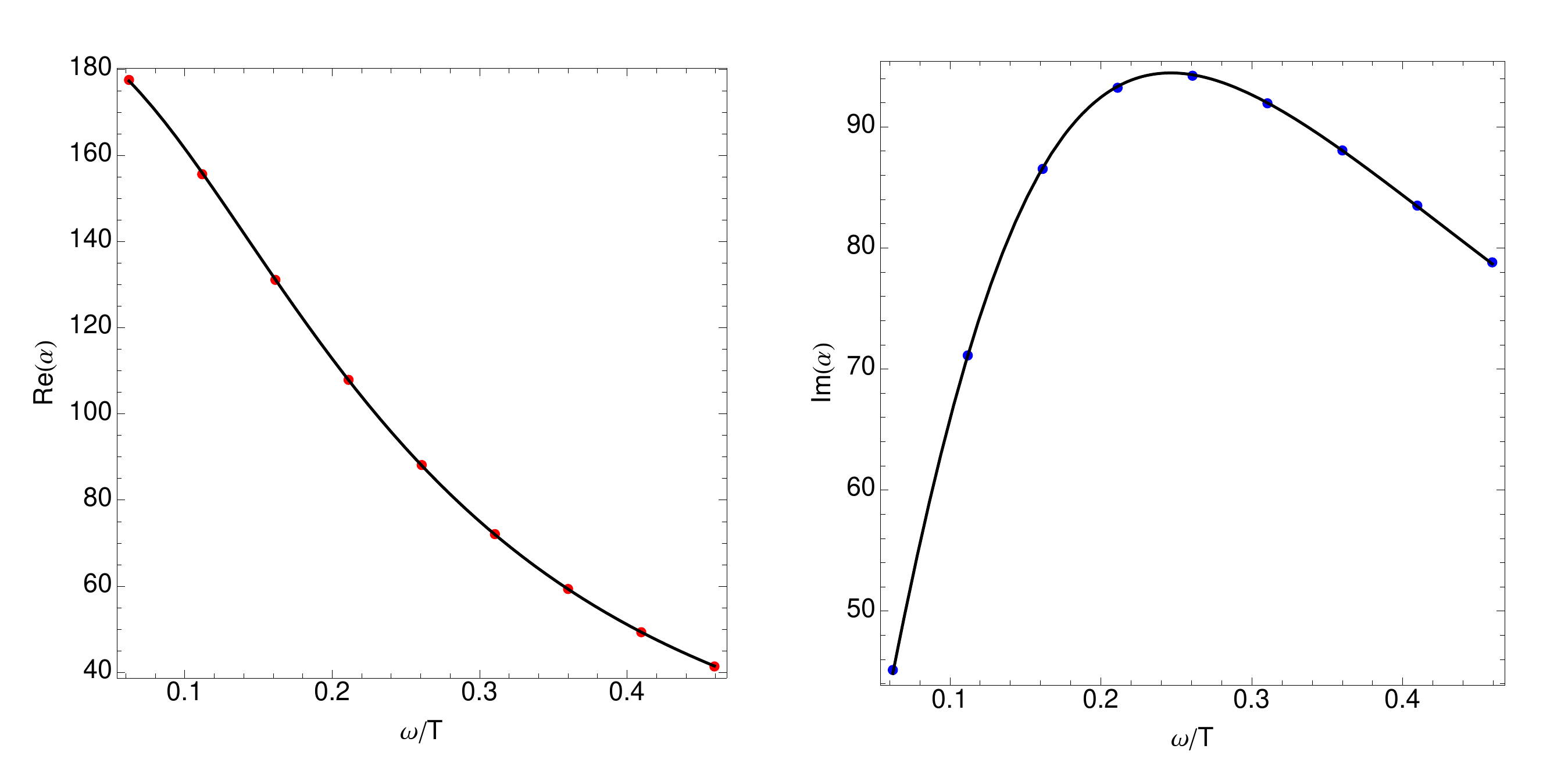}
}
\caption{The real and imaginary parts of the thermoelectric conductivity for small frequencies. The lines are  a fit to the Drude form. This is for $k_0 = 1, A_0 = .75,\mu = 1.4, T/\mu = .115$.}
\label{fig:alphasmall}
\end{figure}

\begin{figure}
\centerline{
\includegraphics[width=0.9\textwidth]{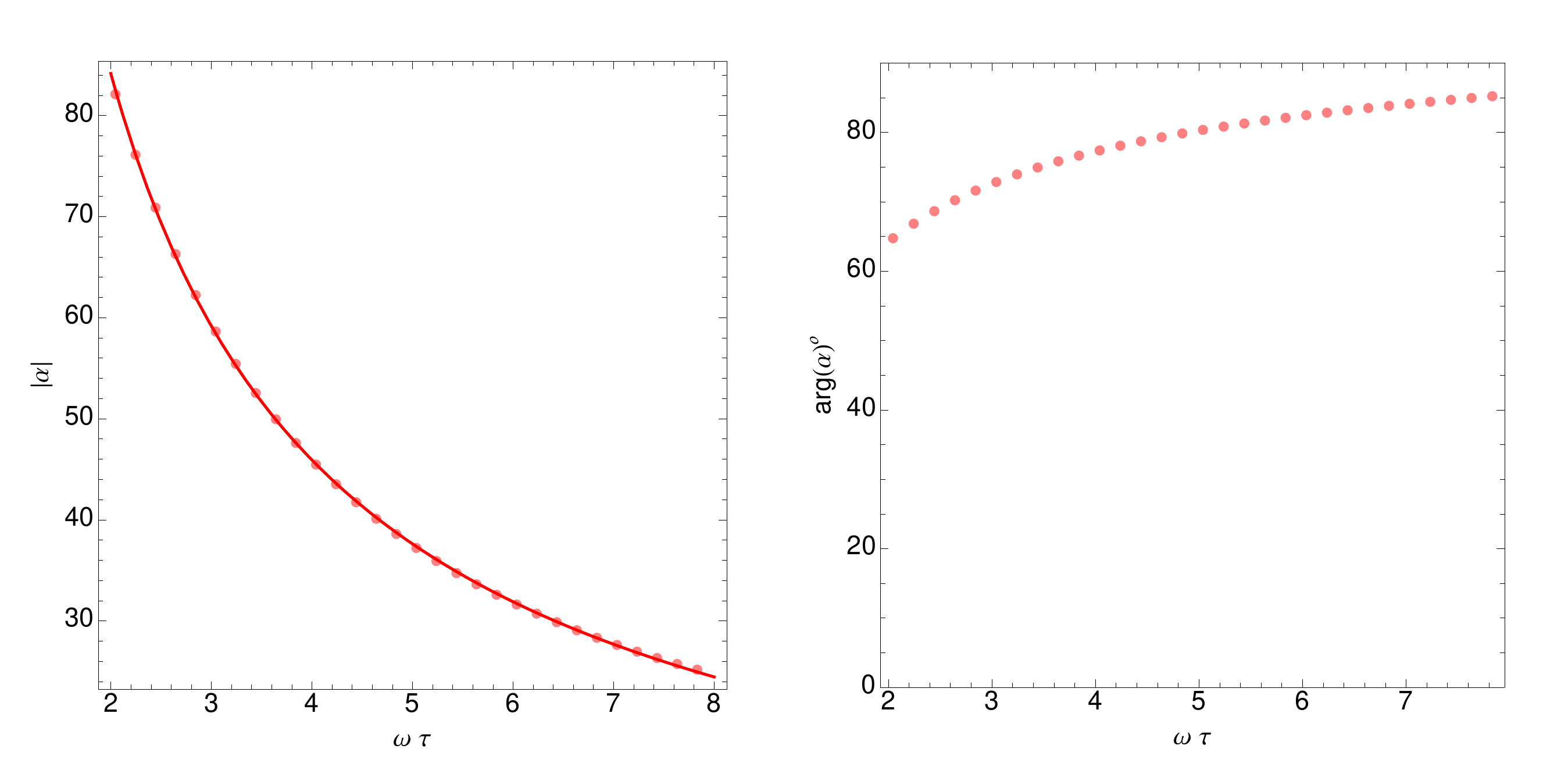}
}
\caption{The magnitude and phase of the thermoelectric conductivity in the intermediate scaling regime. The curve on the left is a fit to the power-law \eqn{newscaling} which determines $\eta \approx 5/6$. Like the previous figure, this is for $k_0 = 1, A_0 = .75,\mu = 1.4, T/\mu = .115$.}
\label{fig:alphapower}
\end{figure}

\begin{itemize}

\item At small frequencies, the thermoelectric conductivity is shown in Fig.~\ref{fig:alphasmall}. It is well approximated by the Drude form, with the scattering time $\tau$ identical (to within a precision of $0.1\%$) to the scattering time extracted from the optical conductivity. 
\item In the intermediate frequency range $2< \omega \tau <8$, where the the optical conductivity follows  a power-law, the thermoelectric conductivity also has power-law form, 
\be |\alpha(\omega)| = \frac{\tilde B}{\omega^{\eta}} + \tilde C\label{newscaling}
\ee
where the exponent is  $\eta \approx 5/6$. This is shown in Fig.~\ref{fig:alphapower}. As far as we can tell, this power is robust against changing the lattice size $k_0$ and the temperature $T$.  
\item At large frequencies, the thermoelectric conductivity approaches
\be
\alpha  = -\frac{ \mu \sigma}{T} + i\frac{\rho}{\omega T}
\ee
which is the result in the absence of a lattice \cite{hh,herzog}. Note that the real part of  $\alpha$ becomes negative at large frequency.
\end{itemize}
 In the absence of a lattice, the three conductivities $\sigma$, $\alpha$ and $\bar \kappa$ are related by Ward identities \cite{hh,herzog}. However, the proof relies on translational invariance and, in the presence of a lattice, the relationship between them involves the background fields. It would clearly be of interest to understand how the exponent $\eta$ for the thermoelectric conductivity is related to the exponent $\gamma$ for the optical conductivity.

\section{A Five-Dimensional Holographic Lattice}\label{sec:5d}

The previous results all hold for asymptotically $AdS_4$ backgrounds. Here we extend our study to asymptotically $AdS_5$ backgrounds, corresponding to $d=3+1$ boundary theories. We adopt the same Einstein-Maxwell-scalar action \eqn{eq:action}, but now in five dimensions with scalar mass $m^2L^2 = -15/4$.
The asymptotic behavior of the scalar is now
\be \Phi = z^{3/2}\phi_1 + z^{5/2}\phi_2 + \cdots
\ee
and we again impose the boundary condition \eqn{eq:source1} on $\phi_1$.
The background holographic lattice can be constructed as before. However, the perturbation to compute the conductivity has an extra complication: we run into the peril of the log \cite{Taylor:2000xw}. Imposing a homogeneous, oscillating electric field at infinity requires the  asymptotic vector potential  to behave like  \cite{garymatt}
\be A_x \rightarrow A_x^{(0)} + A_x^{(2)}(\Lambda_1)\,z^2 -\frac{1}{2}A_x^{(0)}\omega^2\,z^2\log(\Lambda_1 z) + {\cal O}(z^3)\nn\ee
where $\Lambda_1$ is an arbitrary scale. Since changing $\Lambda_1$ changes the finite coefficient of $z^2$, we have indicated that $A_x^{(2)}$ depends on the choice of $\Lambda_1$.  However, the entire expression is independent of  $\Lambda_1$. The bulk Maxwell action also has a log divergence when the Maxwell field is nonzero at infinity. So to obtain a finite action we must add a counterterm
\be
S_{c.t.} = \log(\epsilon \Lambda_2) \int_{z=\epsilon} d^4 x F^2
\ee
where $\epsilon$ is a UV cut-off. The resulting expression for the current, (obtained by varying the finite action with respect to $A_x^{(0)}$) is
\be\label{current}	
J_x = 2A_x^{(2)}(\Lambda_1) -\omega^2 A_x^{(0)}\left[\log\left(\frac{\Lambda_1} {\Lambda_2}\right) - \frac{1}{2}\right]
\ee
This is independent of $\Lambda_1$ but depends on $\Lambda_2$. However since $\Lambda_2$ multiplies the boundary action, this just corresponds to the usual log running of the four dimensional gauge coupling constant. Note that this ambiguity only affects the high frequency part of the current. In terms of calculating the conductivity, the scale at which this log running is important is probably set by the chemical potential $\mu$. Since we are interested in $\omega \ll \mu$, it is unimportant in the region of interest. In the calculations we report below, we have chosen $\Lambda_2$ so that the second term in \eqn{current} (proportional to $\omega^2$) vanishes. We have also verified that other choices of  $\Lambda_2$ with $\log \Lambda_1/\Lambda_2$ as large as ten do not affect our results of how the delta-function is resolved.

A related problem is that a naive application of spectral methods does not yield the required convergence due to the $\log$. This can be overcome by constructing two different discretizations of the bulk spacetime; one close to the UV boundary designed to deal with the log; the other near the black hole horizon. Near the boundary we use a second order finite different scheme, and deep in the bulk we used a Chebyshev grid. They are then patched together. One can then solve for the linearized perturbation and compute the current.

\begin{figure}
\centerline{
\includegraphics[width=0.9\textwidth]{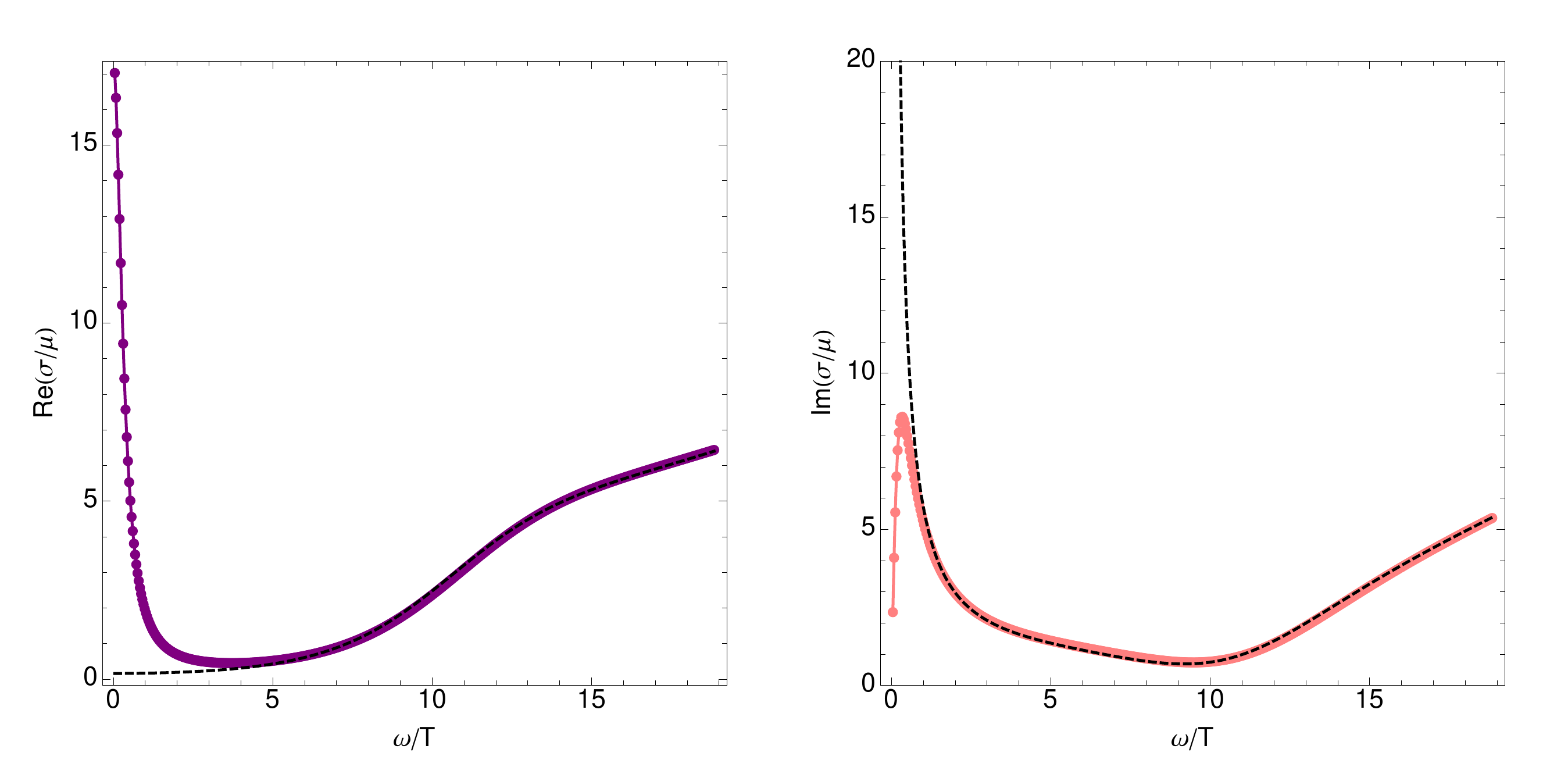}
}
\caption{The optical conductivity both with the lattice (solid line and data points) and without  (dashed line) for the $3+1$ dimensional conductor with  $T/\mu= .21$ and $\mu = 1$. The lattice has $A_0 = 1.5$ and $k_0 = 2$.
 \label{fig:5dlong}}
\end{figure}

\begin{figure}
\centerline{
\includegraphics[width=0.9\textwidth]{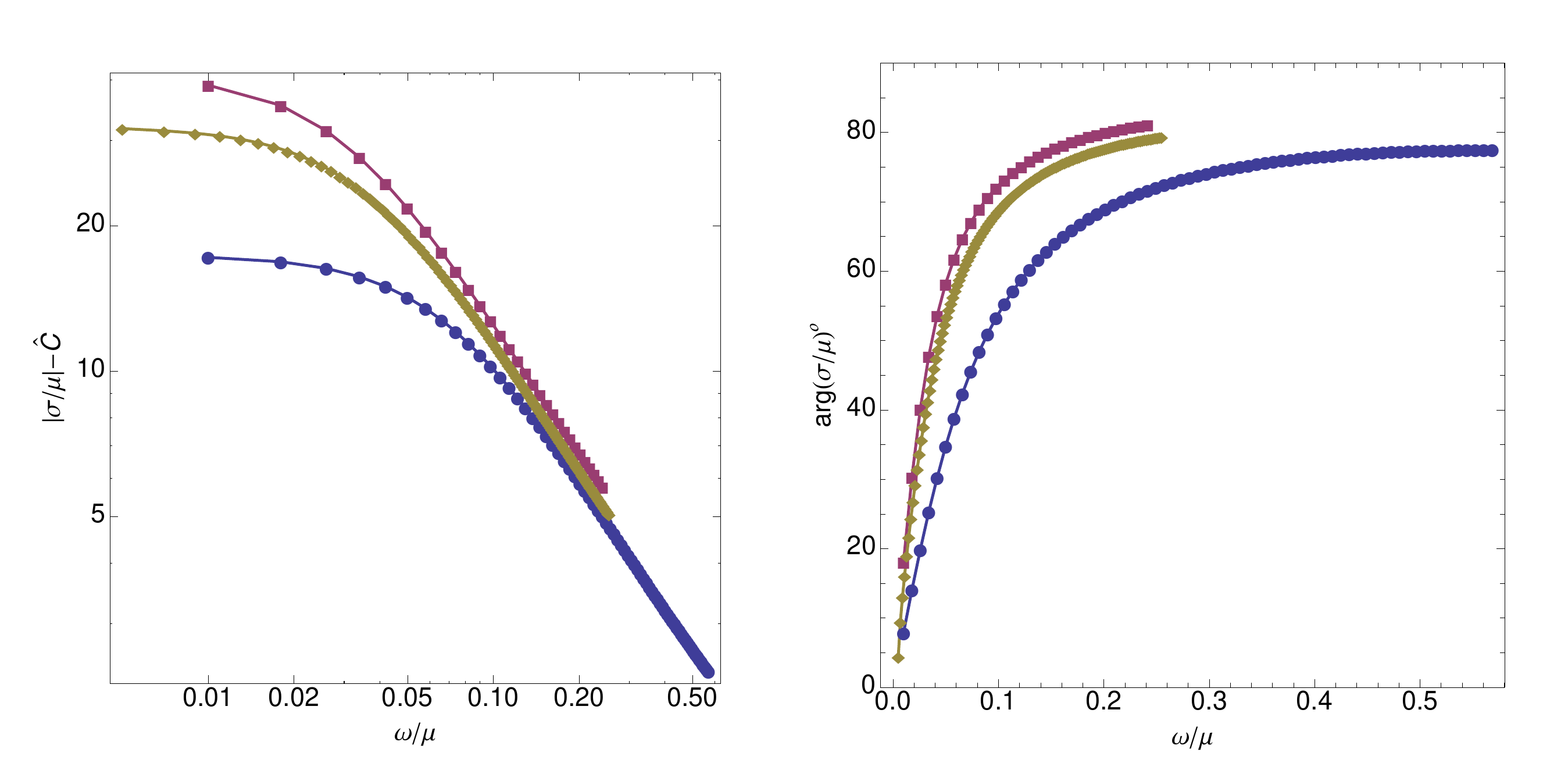}
}
\caption{The optical conductivity for a $3+1$ dimensional system as a function of frequency at three different wavenumbers: squares denote $k_0 = 1$, circles denote $k_0 = 2$, and diamonds denote $k_0 = 3$. The temperature is $T/\mu = .21$, amplitude is $A_0/k_0 = 3/4$, and chemical potential is $\mu = 1$. On the left is a log-log plot of the magnitude of the  conductivity, and  on the right is the phase. \label{fig:5dlogs}}
\end{figure}

The resulting optical conductivity is shown in Fig.~\ref{fig:5dlong}. In $3+1 $ dimensions, the conductivity is no longer dimensionless, so we  plot $\sigma / \mu$.
 A detailed examination  shows results very similar to the  $AdS_4$ case summarized in section 2. At low frequency, the conductivity  follows the simple  Drude form  \eqn{drude}. At intermediate frequency, the conductivity again enters a scaling regime. Only the exponent in the power-law changes. We now find
\be |\sigma(\omega)| = \frac{\hat B}{\omega^{\gamma}}+\hat C\label{morepower}\ee
with $\gamma \approx 0.87 \approx \sqrt{3}/2$.   As in previous examples, this exponent is robust against changes to the parameters in our model. We illustrate this in Fig.~\ref{fig:5dlogs} where we show the conductivity as a function of frequency for three different values of the lattice spacing. On the left is a log-log plot of $|\sigma|-\hat C$,
 and  on the right is the phase. The $k_0 =2$ line extends to larger frequency just because the Drude relaxation time $\tau$ is smaller.  Remarkably, the scaling regime again holds for the same range of frequency (relative to $\tau$) as in the lower dimensional case: $2< \omega \tau < 8$. In the region of the powerlaw, the lines in the log-log plot again lie essentially on top of each other showing that not only is the exponent independent of $k_0$, but the coefficient $\hat B$ depends only weakly on it. For the cases we have checked, the off-set $\hat C$ is about an order of magnitude smaller than the off-set in the $2+1$ theory.

\begin{figure}
\centerline{
\includegraphics[width=0.35\textwidth]{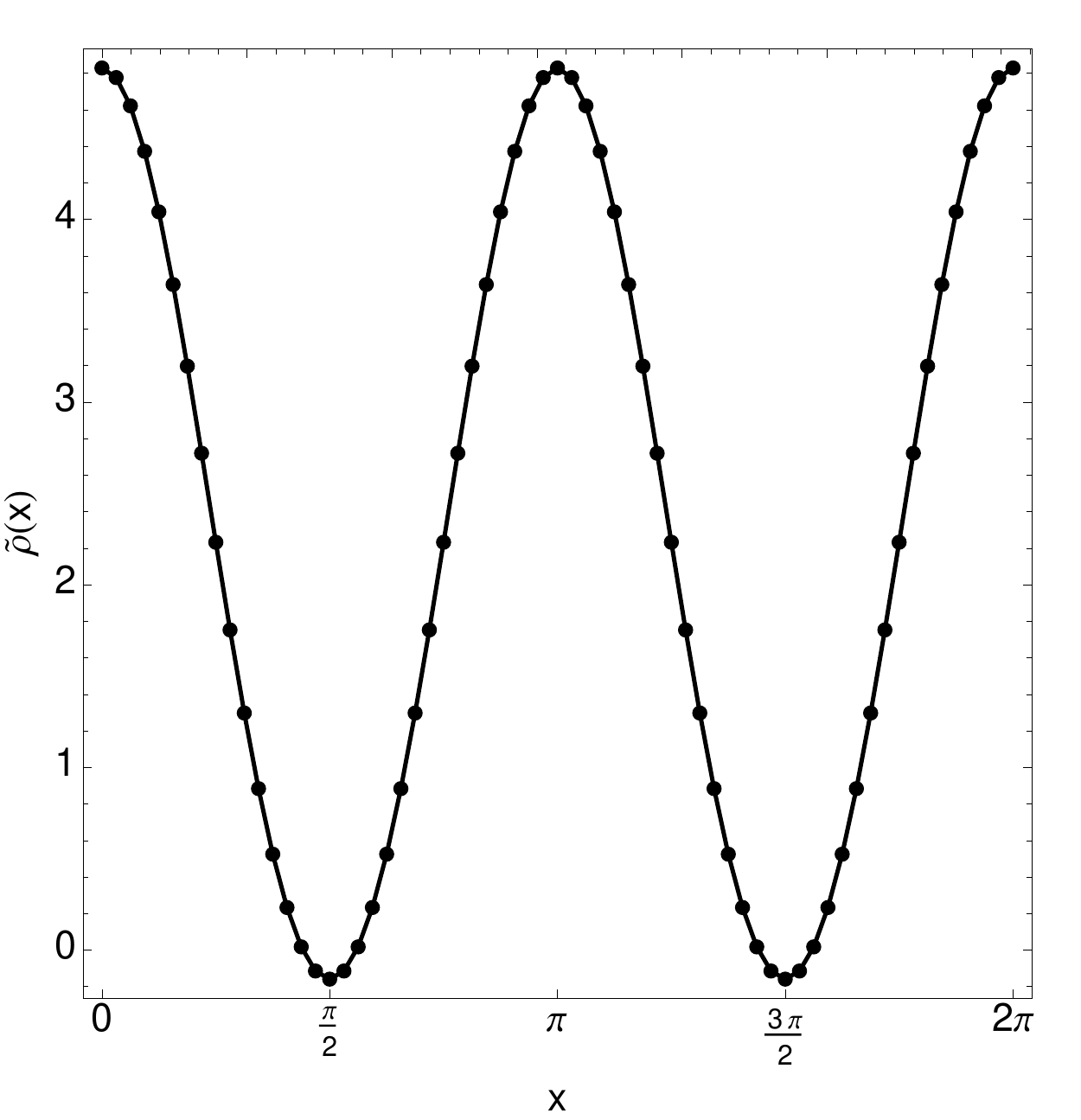}
}
\caption{Charge density variations in the ionic lattice with $k_0 = 2, A_0 = .5, \bar \mu = 2, T/\bar\mu = .055$.}
\label{fig:charge}
\end{figure}

\section{An Ionic Lattice}
We now consider a different kind of lattice. We throw away the scalar field and work with just the Einstein-Maxwell action in four dimensions. 
The lattice is introduced through a spatially varying chemical potential. This type of lattice was used  in several earlier discussions (see, e.g., \cite{seandiego,maeda,maeda2,schalm,Flauger:2010tv,siopsis}) but often in a probe approximation that ignored the backreaction on the metric.
The varying chemical potential can be viewed as representing the potential felt by electrons in an array of ions and, for this reason, the  lattice is often referred to as an ionic lattice.

As usual, the chemical potential is fixed by the boundary value of the temporal component of the gauge field, $A_t$. As $z\rightarrow 0$, we choose
\be  A_t\rightarrow \mu(x) \equiv \bar{\mu} \left[1 + A_0\cos(k_0x)\right]\label{wavy}\ee
Once again we introduce the lattice only in one dimension for computational convenience. The nonlinear PDE's are solved using the same numerical methods as before.

The resulting background solutions are  again parameterized by four variables: the average chemical potential, $\bar{\mu}$, the size of the modulation $A_0$, the lattice wavenumber $k_0$ and the temperature $T$. The resulting physics depends only on three dimensionless quantities which can be taken to be $A_0$, $k_0/\bar{\mu}$,  and $T/\bar{\mu}$.

\begin{figure}
\centerline{
\includegraphics[width=.9\textwidth]{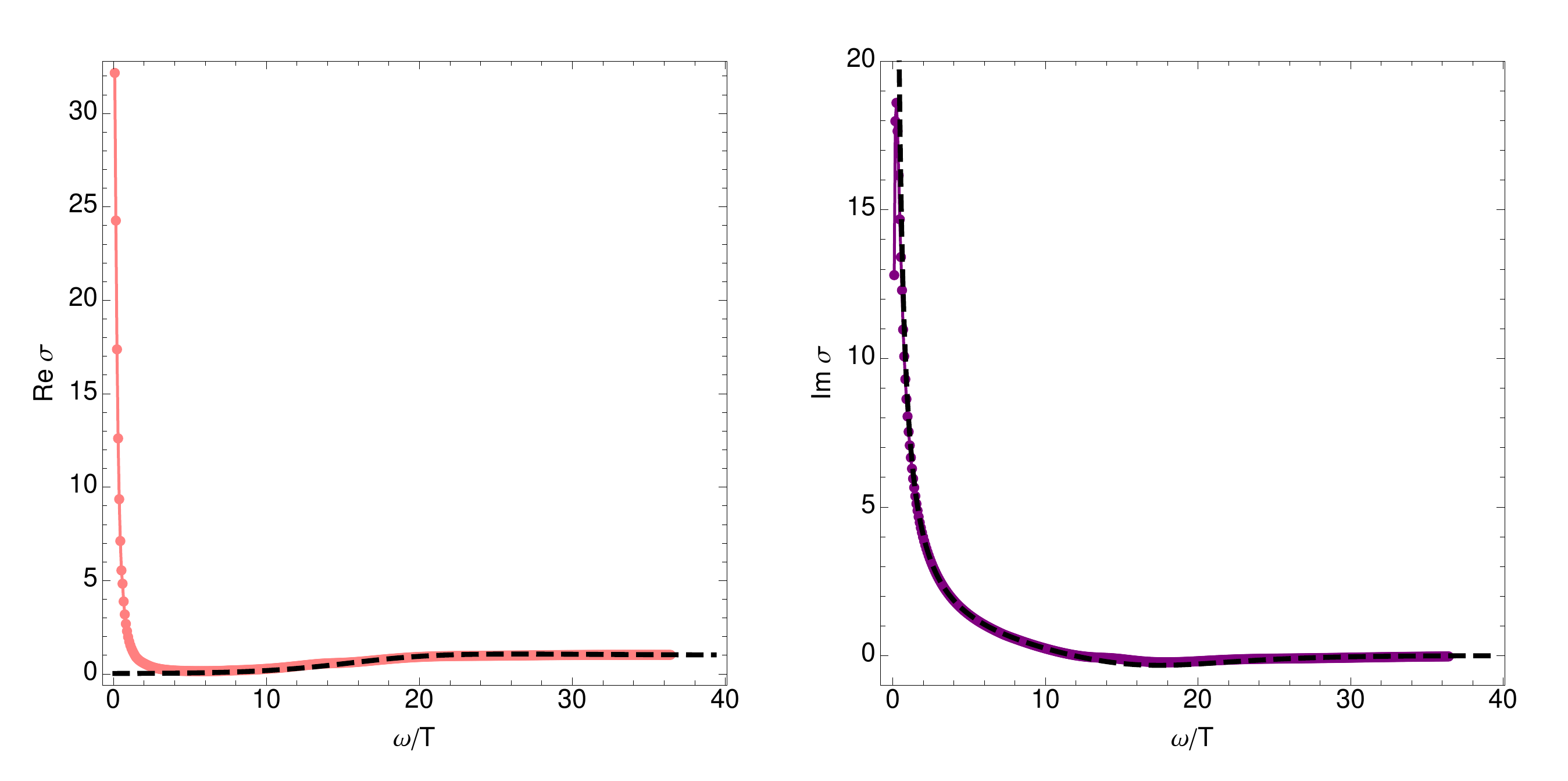}
}
\caption{The optical conductivity both with the ionic lattice (solid line and data points) and without  (dashed line) for the $2+1$ dimensional conductor with $\bar\mu = 2$ and $T/\bar\mu= .055$. The lattice has $A_0 = .5$ and $k_0 = 2$.}
\label{fig:ioniclong}
\end{figure}
\begin{figure}
\centerline{
\includegraphics[width=.9\textwidth]{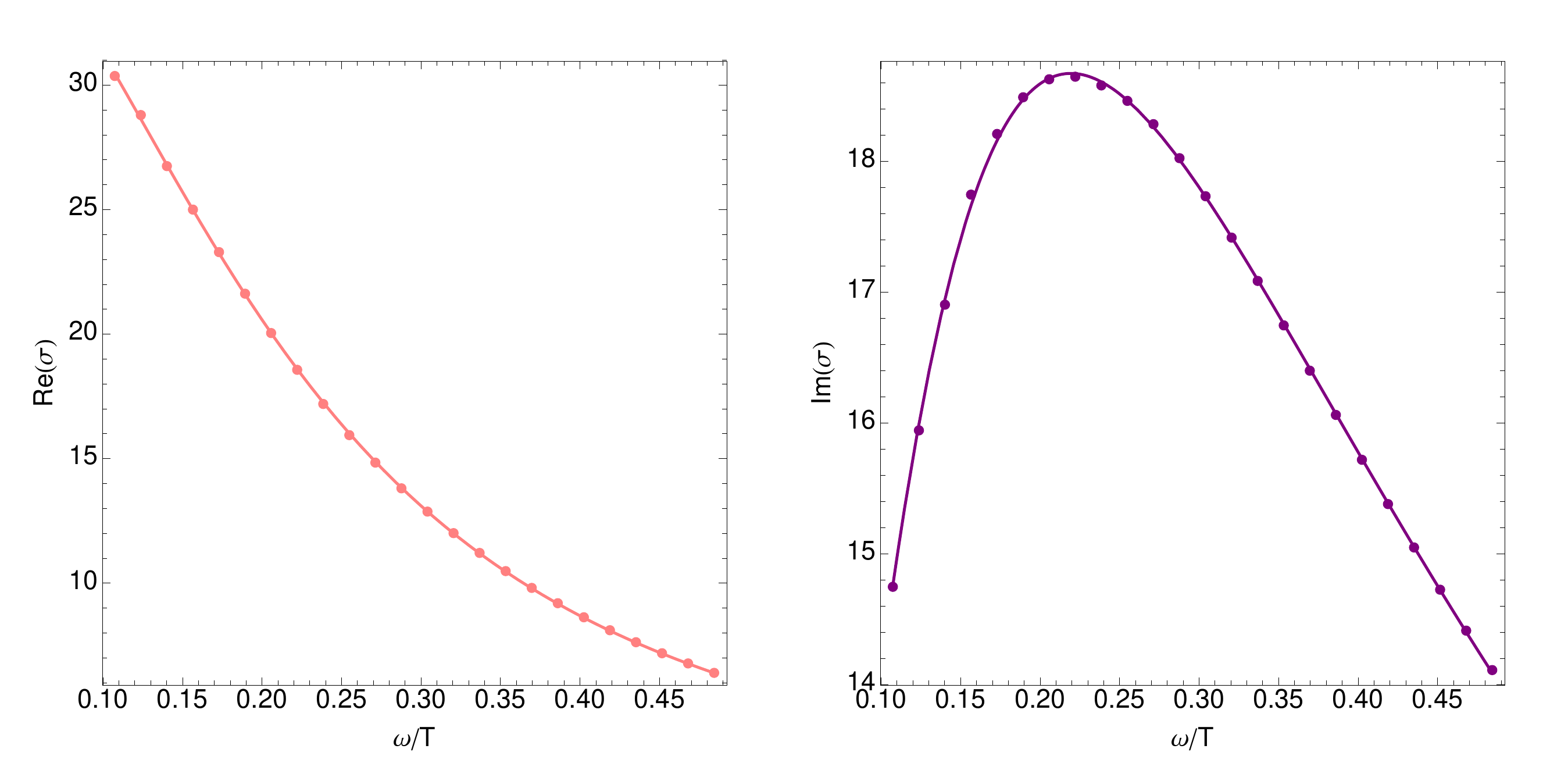}
}
\caption{A blow up of the low frequency regime of Fig.~\ref{fig:ioniclong}. The curve  is a fit to the Drude form \eqn{drude}.}
\label{fig:ionicdrude}
\end{figure}

The charge density, $\tilde \rho(x)$,  of the boundary theory can be read, as always, from the subleading term in the expansion of $A_t$ near the boundary. We plot this in Figure \ref{fig:charge}. Note that the variations in the charge density for the ionic lattice are of order one. This is in contrast to the scalar lattice introduced in \cite{us} where the charge density varied only at the $1\%$ level\footnote{There is one further difference between the two lattices. For the ionic lattice, the modulation in the charge density has wavenumber $k_0$; for the scalar lattice, the backreaction on both the metric and gauge field is proportional to $\phi^2$ which ensures that if $\phi$ varies with wavenumber $k_0$, the charge density varies with wavenumber $2k_0$.}.

To compute the optical conductivity, we next perturb the background, fixing the usual boundary condition on $A_x$ \eqn{Abdycond}. The calculation is similar to that in \cite{us} except that there is one less function to determine because there is no scalar field to perturb. The conductivity is again given by \eqn{eq:opticalderiv}. One might worry that this expression is no longer gauge invariant since the ionic lattice has an asymptotic background electric field in the $x$ direction, so $\delta F_{xt}$ is not invariant under diffeomorphisms.  However, gauge transformations only consist of diffeos that are the identity at infinity, so the asymptotic value of $\delta F_{xt}$ (which is the only thing that enters the conductivity) is still gauge invariant.

The resulting optical conductivity is shown in Fig.~\ref{fig:ioniclong} for a wide range of frequencies. For comparison, the optical conductivity without the lattice is shown as a dashed line. Once again we see that the zero frequency delta function is smeared out. A blow up of the low frequency regime is shown in Fig.~\ref{fig:ionicdrude}. The data is again well fit by the simple two parameter Drude form \eqn{drude}.  The constant $K$ agrees with the coefficient of the pole in Im~$\sigma$ in the translationally invariant case, so it is really just a one parameter fit.

\begin{figure}
\centerline{
\includegraphics[width=.9\textwidth]{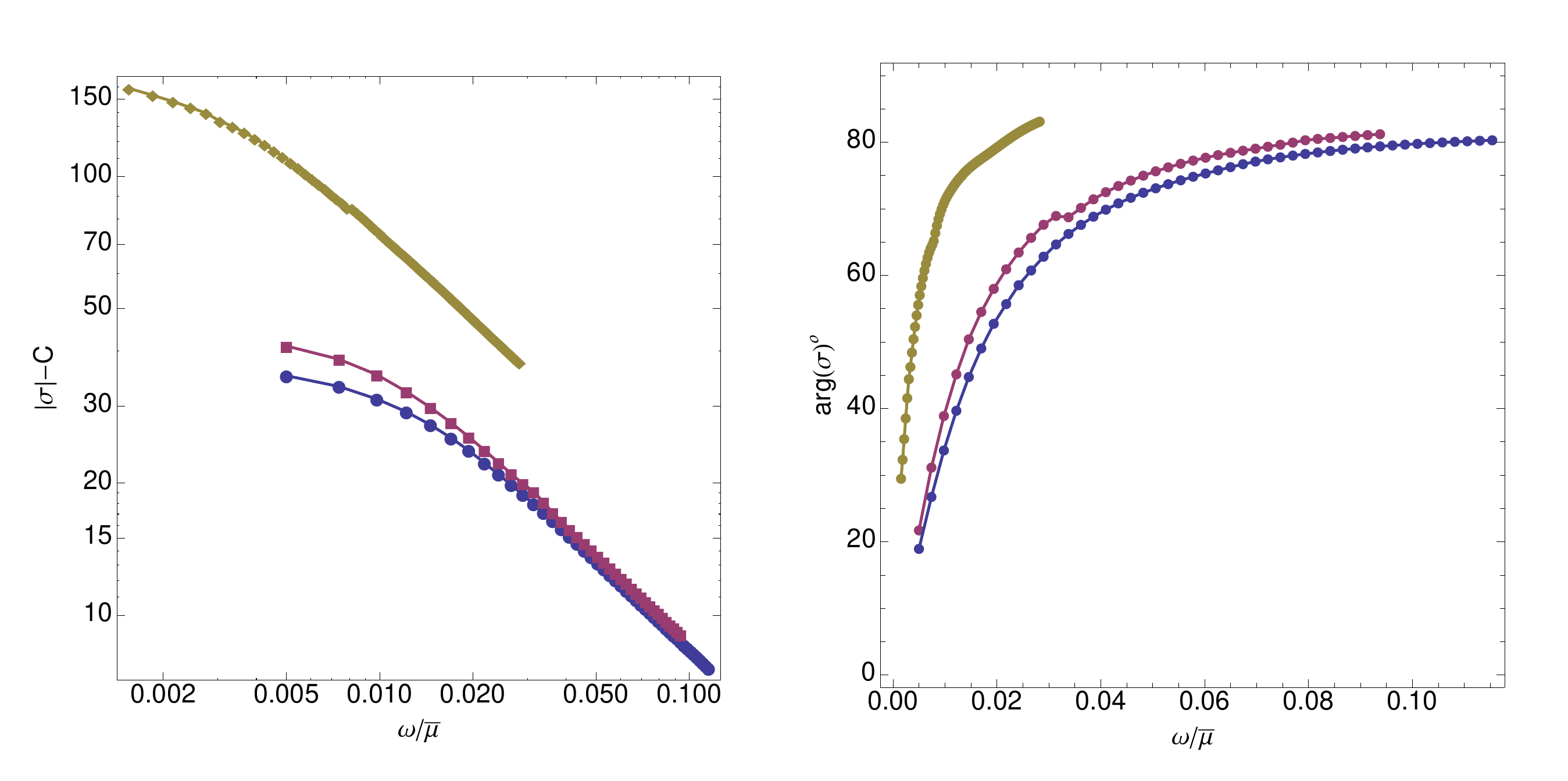}
}
\caption{The optical conductivity for a $2+1$ ionic lattice as a function of frequency at three different wavenumbers: squares denote $k_0 = 1$, circles denote $k_0 = 2$, and diamonds denote $k_0 = 3$. The temperature is $T/\bar\mu = .055$, chemical potential $\bar\mu = 2$, and amplitude $A_0 = .5$. On the left is a log-log plot of the magnitude of the  conductivity, and  on the right is the phase.}
\label{fig:ionicseveral_k0}
\end{figure}

\begin{figure}
\centerline{
\includegraphics[width=.9\textwidth]{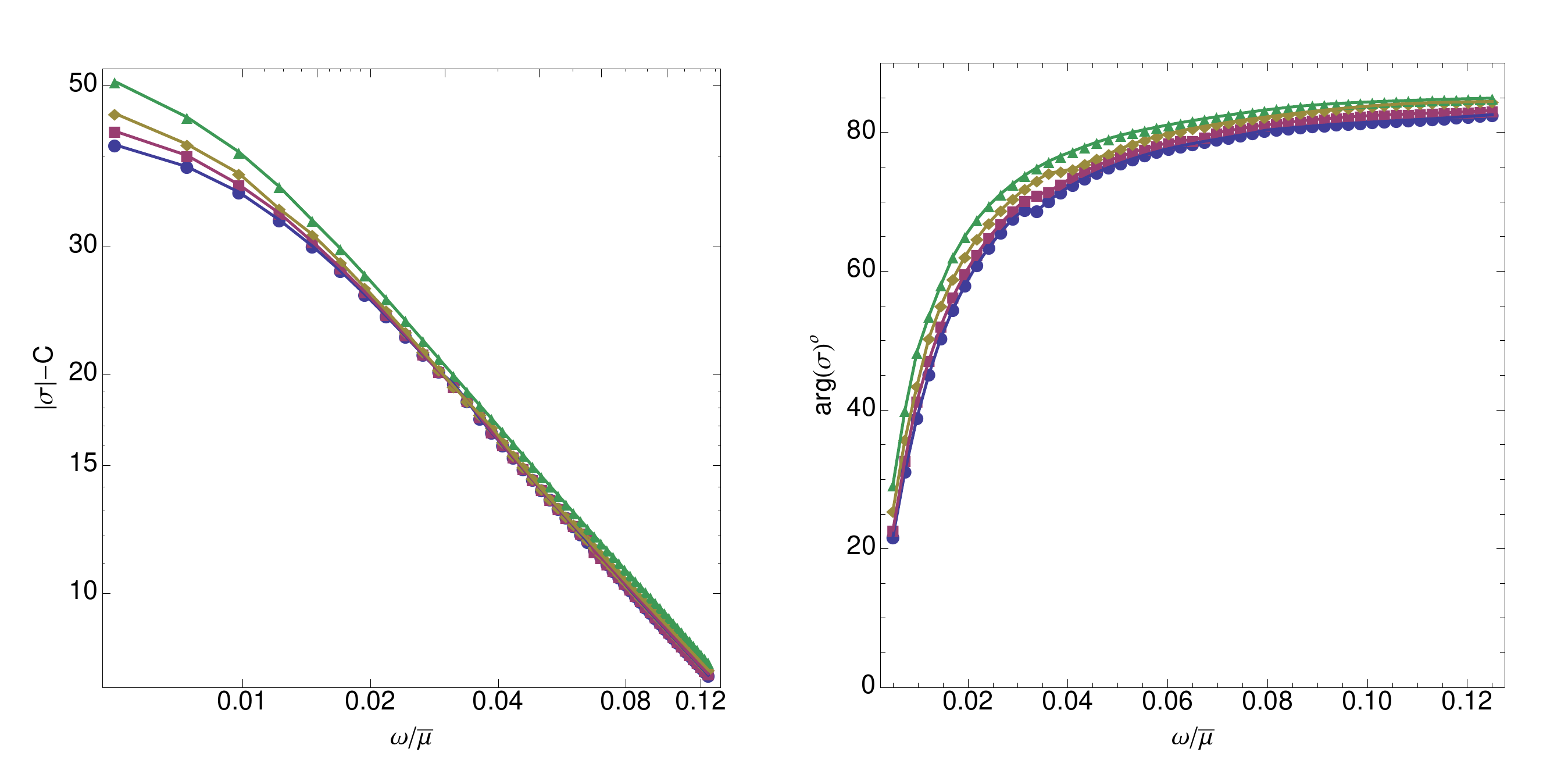}
}
\caption{The optical conductivity for a $2+1$ ionic lattice as a function of frequency at four different temperatures: circles denote $T/\bar\mu =.055 $, squares denote $T/\bar\mu = .049$,  diamonds denote $T/\bar\mu =.041 $, and triangles denote $T/\bar\mu =.033 $. The lattice has $A_0 = .5$ and $k_0 = 2$. On the left is a log-log plot of the magnitude of the  conductivity, and  on the right is the phase. }
\label{fig:ionicseveral_t}
\end{figure}

 At slightly larger frequencies, the absolute value of the conductivity is well fit by the power law behavior \eqn{power} with the same exponent $\gamma = 2/3$. It again holds in the same range of frequency (relative to the relaxation time found in the Drude fit) as our earlier lattice: $2 < \omega \tau < 8$.  The exponent $2/3$ is independent of the lattice spacing and temperature as shown in Fig.~\ref{fig:ionicseveral_k0} and Fig.~\ref{fig:ionicseveral_t}. The coefficient $B$ is again essentially independent of temperature, but shows some dependence on lattice spacing. The similarity between Fig.~\ref{fig:ionicseveral_t} and the data shown in \cite{vandermarel} seems remarkable.
 
 The fact that the same scaling behavior is seen for two different lattices shows that it is a property of the underlying system rather than the detailed form of the lattice. 

\subsection{Resonances}

\begin{figure}
\centerline{
\includegraphics[width=.9\textwidth]{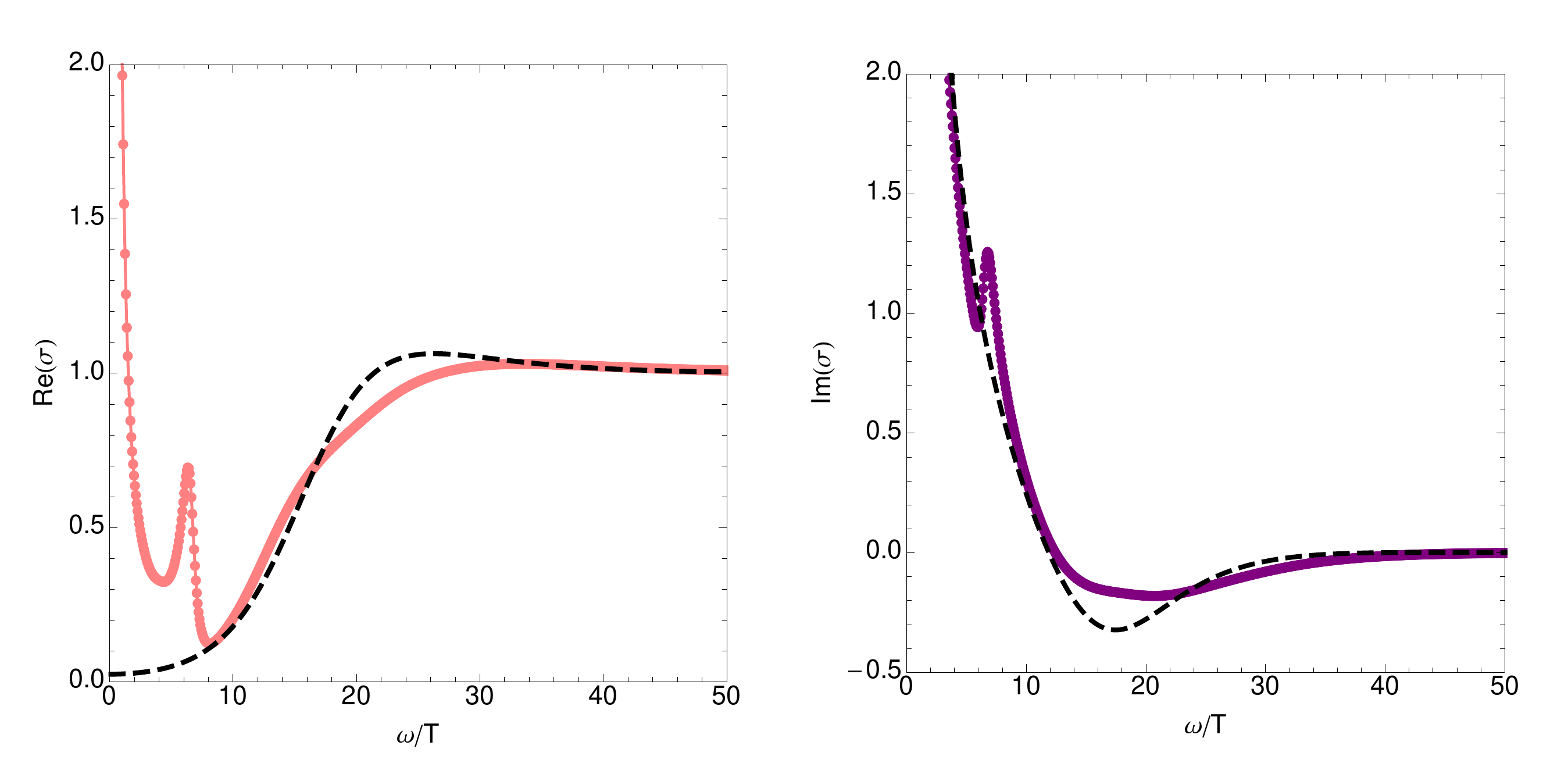}
}
\caption{The optical conductivity for a lattice with the same parameters as Fig.~\ref{fig:ioniclong} except that now $k_0 =1$. We have greatly expanded the vertical scale. }
\label{fig:surprise1}
\end{figure}

There is one further feature in the optical conductivity which is not present in the homogeneous case: resonances. A particularly clean example is shown in Figure \ref{fig:surprise1} (plotted with the same parameters as Figure \ref{fig:ioniclong} but with lattice spacing $k_0=1$). Such resonances are common place in condensed matter systems where they arise from bosonic quasi-particles carrying a dipole moment such as optical phonons or excitons. In particular, they are often seen in the cuprates \cite{Takenaka}. In the present case, the resonance can be attributed 
to a quasinormal mode of the striped black hole.   In the homogeneous case, one does not see the effect of quasinormal modes since the perturbation needed to compute the conductivity decouples from them.  This is not true for the lattice background.  Since all the modes couple together, imposing a homogeneous asymptotic electric field will also excite quasinormal modes with wavenumber $k$ equal to integer multiples of the lattice wavenumber $k_0$. Only modes with $k=0$ will contribute to the (homogeneous) conductivity $\sigma(\omega)$. Similar resonances  were recently used as a method to compute black hole quasinormal modes \cite{Berti:2009wx}.

\begin{figure}
\centerline{
\includegraphics[width=.9\textwidth]{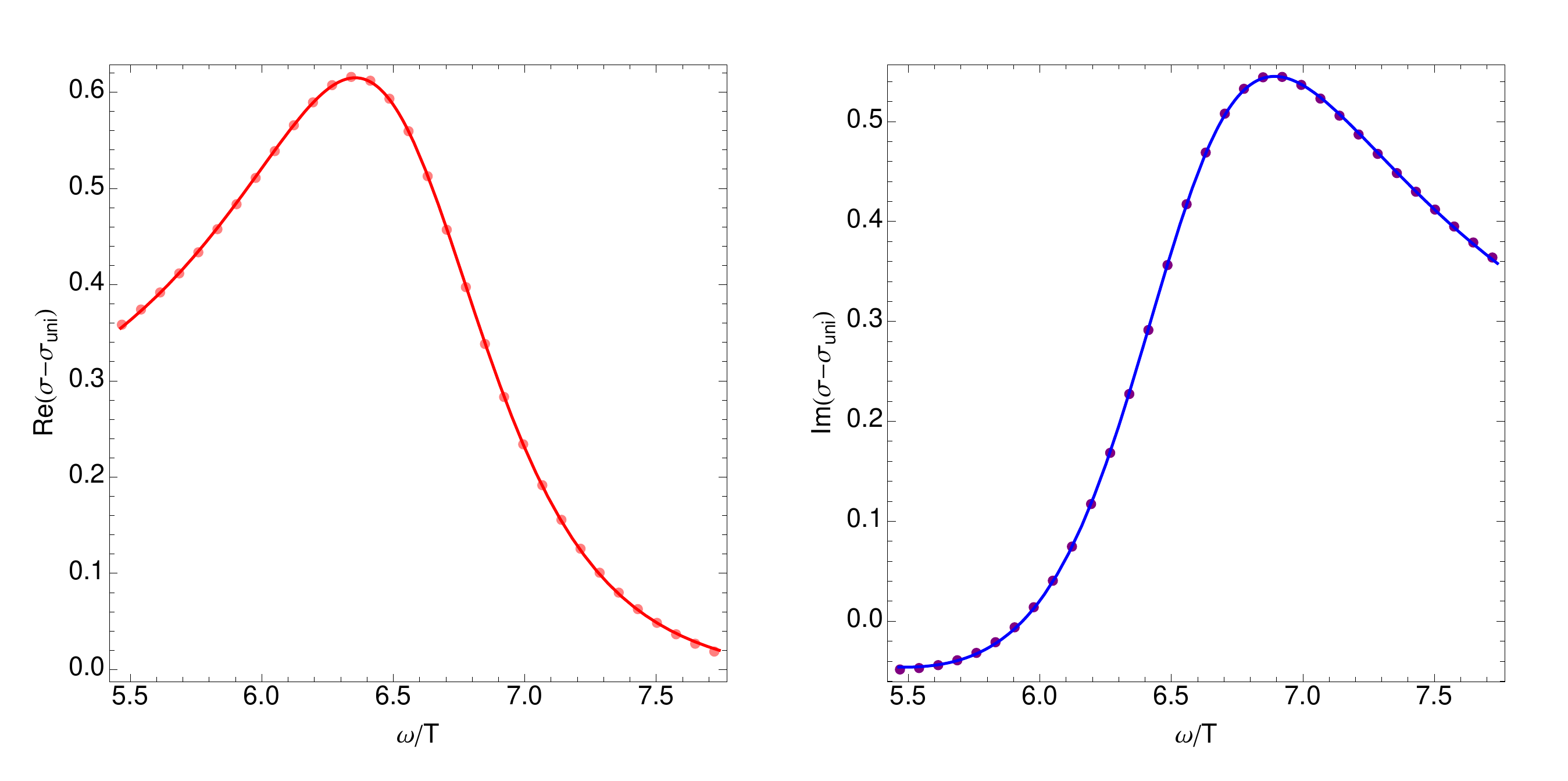}
}
\caption{The optical conductivity near the resonance with the uniform background subtracted off. The curve is a fit to \eqn{qnm}.}
\label{fig:qnm}
\end{figure}

To see the structure of the resonance more clearly, we subtract off the homogeneous background. Fig.~\ref{fig:qnm} shows the result. The data is very well fit by assuming that the retarded Greens function which  determines the conductivity has a pole at a complex frequency $\omega_0$:
\be\label{qnm}
\sigma(\omega) = \frac{G^R(\omega)}{i\omega}= \frac{1}{i\omega} \frac{a+b(\omega - \omega_{0})}{\omega - \omega_{0}}
\ee
where $a$ and $b$ are complex constants.  From the fit, we determine the quasinormal mode frequency to be $\omega_0/T = 6.6 - 0.64i$.

Resonances of this type in the optical conductivity are a generic feature of holographic lattices. They were not seen in \cite{us} either because the lattice (in the metric) had much smaller amplitude, or because the temperature was not low enough.  They are more pronounced in the  ionic lattice, although their precise position depends on the various parameters. For example, the lattice with $k_0=2$ shows only a hint of a bump in  $\sigma(\omega)$ because the real part of the quasinormal mode frequency moves to larger values  where the conductivity is  already approaching its asymptotic plateaux.

We stress that these resonances are independent from the scaling phenomenon which is the primary focus of this paper. Indeed, the eagle-eyed reader can see a remnant of a resonance depicted as  a slight kink in the upper-graph of the log-log plot shown in the left-hand side of Figure \ref{fig:ionicseveral_k0}.

\section{Discussion}

We have extended our earlier study of transport in holographic lattices \cite{us} and presented further evidence that these lattices induce an intermediate scaling regime. The agreement with measurements of the optical conductivity of the cuprates is stronger than we realized earlier, since not only is the $2/3$ exponent in the power law independent of all parameters in our model,  but the coefficient $B$ is temperature independent.  We have seen that $B$ does depend on the lattice spacing, but since changing $k_0$ changes the charge per unit cell, this  is perhaps analogous to doping the material. 

We end with a few comments and speculations on these results.
First,  since we are studying homogeneous transport (at momentum $k=0$), one might imagine that there is an intermediate, homogeneous scaling regime in the geometry. In other words,  is it possible that for some range of radii,  a homogeneous approximation to our lattice geometry has a scaling symmetry such as the solutions in \cite{kiritsis,Iizuka:2011hg}? Unfortunately, it is difficult to test this. To construct a homogeneous approximation to the geometry, one first needs to define slices of constant radial distance and, with all metric functions depending on both $z$ and $x$, it is not obvious how to do this in a diffeomorphism invariant manner. 
Even if such a homogeneous geometry did exist at some intermediate energy regime, it can never truly capture the umklapp physics at play in resolving the delta-function. 

Secondly, in all the cases that we have studied, the scaling regime corresponds to approximately $2 <\omega \tau<8$, where $\tau$ is the relaxation time coming from the Drude fit at low frequency. It is striking that, while $\tau$ differs wildly from model to model, the scaling regime remains constant when parameterized in this fashion. However, it is also clear that the regime is limited and it is not obvious how to extend it in a parametric manner. Our claims that we are seeing scaling rather than, say, a cross-over rely on the robustness of the power-law as all other parameters are varied as evidenced in the plethora of log-log plots that fill this paper. 

We have no analytic understanding of why the regime is always $ 2 <\omega \tau<8$ and it would clearly be of interest to try to extend this range.  One idea is to  try to subtract the homogeneous background conductivity from the lattice conductivity and study the absolute value of the difference. Unfortunately, this does not work; as we have seen,  the lattice conductivity includes contributions from resonances  that are not present in the homogeneous case.

Finally, we have said repeatedly that the exponent in our power-law fall-off is independent of the parameters in our model: the wavenumber $k_0$, temperature $T$, and amplitude $A_0$. However, there are limits of each of these parameters where the effects of the lattice should go away. As either $k_0\rightarrow 0$ or $A_0\rightarrow 0$, the lattice disappears. In the limit $T\rightarrow 0$, the lattice survives but there is no dissipation, and a perfect lattice without dissipation still has infinite DC conductivity. In all these cases, the delta-function at $\omega=0$ must re-emerge. How does this happen? We expect that the mechanism is the same in all limits. The power-law holds only in the range $2 < \omega \tau < 8$ and, in each of these limits,  the scattering time diverges: $\tau\rightarrow \infty$.

\vskip 2cm
\centerline{\bf Acknowledgements}
\vskip 1cm

We are grateful to Subir Sachdev for a number of useful comments on the results of \cite{us}. 
This work was supported in part by the National Science Foundation under Grant No. PHY12-05500, and ERC STG grant 279943, ``Strongly Coupled Systems".

\end{document}